\documentclass[
 reprint,
superscriptaddress,
longbibliography,
 amsmath,amssymb,
 aps,
prx,
]{revtex4-2}

\usepackage{graphicx}
\usepackage{dcolumn}
\usepackage{upgreek}
\usepackage{bm}

\usepackage[version=4]{mhchem}
\usepackage{hyperref}
\hypersetup{
    colorlinks = true,
    citecolor = {blue},
    linkcolor = {blue},
    urlcolor = {black}
}

\begin{document}
\title{Thermoelectric transport in \ce{Ru2TiSi} full-Heusler compounds}

\author{Fabian Garmroudi}
\email{f.garmroudi@gmx.at}
\affiliation{Institute of Solid State Physics, TU Wien, 1040 Vienna, Austria}  
\author{Michael Parzer}
\affiliation{Institute of Solid State Physics, TU Wien, 1040 Vienna, Austria}
\author{Takao Mori}
\affiliation{International~Center~for~Materials~Nanoarchitectonics~(WPI-MANA),~National~Institute~for~Materials~Science,~Tsukuba~305-0044,~Japan}
\affiliation{University of Tsukuba, Tsukuba 305-8577, Japan}
\author{Andrej Pustogow}
\affiliation{Institute of Solid State Physics, TU Wien, 1040 Vienna, Austria}
\author{Ernst Bauer}
\affiliation{Institute of Solid State Physics, TU Wien, 1040 Vienna, Austria}

\begin{abstract}
Heusler compounds with six valence electrons per atom have attracted interest as thermoelectric materials owing to their semimetallic and semiconducting properties. Here, we theoretically and experimentally investigate electronic transport in \ce{Ru2TiSi}-based full-Heuslers. We show that electronic transport in this system can be well captured by a two-parabolic band model. The larger band gap of \ce{Ru2TiSi} promises a higher thermoelectric performance, compared to its isovalent family member \ce{Fe2VAl}, which has been studied as a thermoelectric material for over two decades. Additionally, we identify \textit{p}-type \ce{Ru2TiSi} as far more efficient than previously studied \textit{n}-type compounds and demonstrate that this can be traced back to much lighter and more mobile holes originating from dispersive valence bands. Our findings demonstrate that an exceptionally high dimensionless figure of merit $zT > 1$ can be realized in these $p$-type compounds around 700\,K upon proper reduction of the lattice thermal conductivity, \textit{e.g.}, by substituting Zr or Hf for Ti.
\end{abstract}

\maketitle
\section{Introduction}
Heusler compounds represent a highly tunable material platform encompassing over thousand different members that exhibit a variety of interesting electronic phases, ranging from half-metallicity to semiconducting states and nontrivial topological band structures \cite{graf2011simple,casper2012half,manna2018heusler}. Heusler compounds are typically subcategorized into half-Heuslers (hHs) and full-Heuslers (fHs) with $XYZ$ and $X_2YZ$ stoichiometries, respectively. In half-Heuslers, crystallizing in the non-centrosymmetric C1$_b$ structure (space group no.\,216, F$\overline{4}$3\textit{m}), one of the $X$ sublattices is vacant, whereas the full-Heusler structure with $L1_2$ ordering (space group no.\,225, F\textit{m}$\overline{3}$\textit{m}) can be interpreted as four interpenetrating \textit{fcc} sublattices, two of which are comprised of $X$ atoms, while the others are built-up from the $Y$ and $Z$ atoms. $X$ and $Y$ atoms are usually transition metals, whereas $Z$ is typically a III$^\text{rd}$, IV$^\text{th}$ or even V$^\text{th}$ main group element \cite{graf2011simple}.

In general, research on Heusler materials is guided by simple electron counting rules, like the Slater-Pauling rule \cite{galanakis2002slater,galanakis2006electronic,
graf2011simple,skaftouros2013generalized}, which states that Heusler compounds with an average valence electron concentration (VEC) of 6 valence electrons per atom are non-magnetic semiconductors or semimetals, depending on the strength of hybridization. However, as the VEC increases or decreases, magnetic metals emerge. Although there are few exceptions to this \cite{parzer2022high,parzer2024semiconducting}, hundreds of hH and fH compounds align with this concept and researchers have adhered to these straightforward electron counting rules, as they enable an accurate and quick estimate of the electronic and magnetic ground state properties and serve as fundamental principles \cite{graf2011simple}, particularly in the quest for thermoelectric semiconductors \cite{zeier2016engineering,anand2019double,
wang2022discovery,zhang2022designing}.

The performance of thermoelectric materials is evaluated by the dimensionless figure of merit $zT=S^2\sigma\kappa^{-1}T$, which depends on the Seebeck coefficient $S$, the electrical conductivity $\sigma$ the thermal conductivity $\kappa$ and the absolute temperature $T$, and directly determines the efficiency of a thermoelectric conversion device. For practical purposes, $zT>1$ is commonly considered as a threshold \cite{snyder2008complex}. While numerous hH compounds, \textit{e.g.} (Ti,Zr,Hf)NiSn, (Nb,Ta)FeSb, etc., are already widely recognized as efficient thermoelectric materials \cite{zhu2015high,rogl2023development,li2024half}, their fH relatives have not yet reached $zT$ values, that would make them competitive with state-of-the-art materials \cite{pecunia2023roadmap}. 

\begin{figure}[b!]
\newcommand{\setwidth}{0.45}
			\centering
			\hspace*{-0.0cm}
		\includegraphics[width=0.45\textwidth]{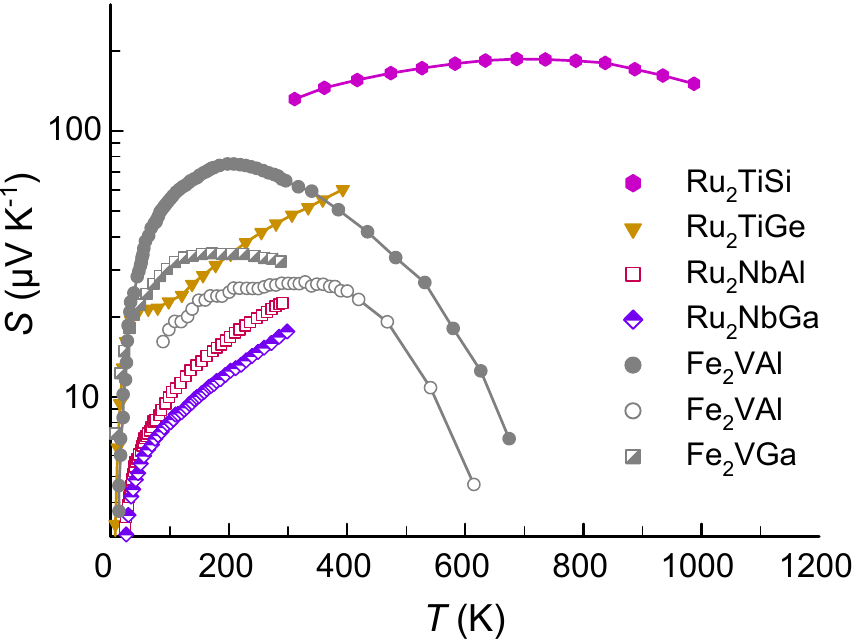}
	\caption{Comparison of temperature-dependent Seebeck coefficient of various semimetallic and semiconducting full-Heuslers with six valence electrons per atom (VEC\,=\,6).} 
	\label{Fig1}
\end{figure}

\begin{figure*}[tbh]
\newcommand{\setwidth}{0.45}
			\centering
			\hspace*{0cm}
		\includegraphics[width=0.8\textwidth]{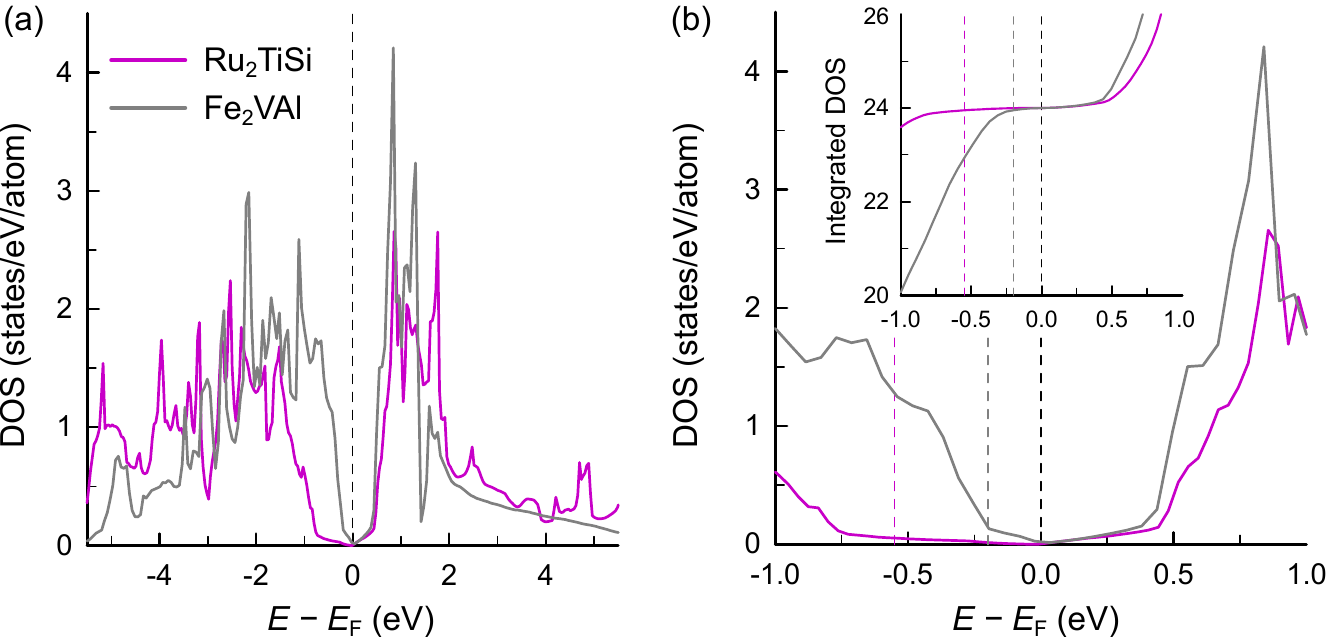}
	\caption{\textbf{Electronic density of states of the full-Heusler compounds \ce{Ru2TiSi} and \ce{Fe2VAl}}. (a) A deep pseudogap with an almost negligibly small DOS at the Fermi level is present in both compounds, which is framed by sharp features in the DOS, originating from rather localized Fe-$3d$ and Ru-$4d$ states. (b) A close-up around the Fermi energy emphasizes that the pseudogap is much broader in \ce{Ru2TiSi}. Additionally, it can be seen that the dispersive pseudogap states, reaching into the gap region (small DOS), especially the valence band states, have a much larger bandwidth and are therefore even more dispersive for \ce{Ru2TiSi} compared to \ce{Fe2VAl}. Black dashed line indicates the Fermi energy $E_\text{F}$, and grey and purple dashed lines show $E_\text{F}$ for a rigid-band doping scenario with 0.05\,holes per formula unit. Inset shows the integrated DOS of both compounds.} 
	\label{Fig2}
\end{figure*}

\ce{Fe2VAl} is unarguably the most prominent fH thermoelectric material with attractive near-room temperature thermoelectric properties \cite{nishino2006thermal,garmroudi2022large,alleno2023optimization}. Despite consisting only of metallic elements, \ce{Fe2VAl} is a semimetal with a tiny band overlap or an almost gapless semiconductor \cite{okamura2000pseudogap,anand2020thermoelectric,hinterleitner2021electronic}. There are two primary factors limiting the performance of \ce{Fe2VAl} thermoelectrics: (i) Their intrinsically large lattice thermal conductivity and (ii) the small band gap, which results in bipolar conduction already at $T \lesssim 300\,$K. Fig.\,\ref{Fig1} compares the temperature-dependent Seebeck coefficient $S(T)$ of various fH compounds with $\text{VEC}=6$, reported and experimentally studied in the literature previously \cite{lue2002thermoelectric,knapp2017impurity,
mondal2018ferromagnetically,mondal2023thermoelectric,
fujimoto2023enhanced}. Note the pronounced maximum in $S(T)$ of \ce{Fe2VAl} at around 200\,K, arising from the aforementioned bipolar conduction, \textit{i.e.}, the activation of minority carriers across the narrow band gap. On the other hand, isovalent \ce{Ru2TiSi}, a novel fH compound recently studied by Fujimoto \textit{et al.}\,\cite{fujimoto2023enhanced}, displays a much larger Seebeck coefficient and broad maximum, vastly surpassing all other known fH compounds over the entire temperature range. This motivated us to experimentally and theoretically study in detail the electronic transport in \ce{Ru2TiSi} and assess its potential thermoelectric performance at optimized doping. 
\section{Experimental Methods and Modelling}
Polycrystalline \ce{Ru2TiSi_{1-x}Al_x} materials were synthesized by melting raw elements with high purity (99.99\,\% Ru, 99.95\,\% Ti, 99.9999\,\% Si and 99.999\,\% Al) using a high-frequency induction heating furnace. Even though powder X-ray diffraction investigations displayed a single Heusler phase directly after the melting procedure, the samples were further annealed at 1273\,K for two days in vacuum (10$^{-5}$\,mbar) to optimize homogeneity. The samples were then cut using a high-speed cutting device (Accutom by the company Struers) equipped with a diamond cutting wheel. The electrical resistivity and Seebeck coefficient at high temperatures were measured in the temperature range 300\,--\,860\,K in an inert He atmosphere using a commercially available setup (ZEM3 by ULVAC-RIKO). To analyze the temperature- and carrier concentration-dependent Seebeck coefficient, we employed a least squares fit model based on Boltzmann transport theory and the parabolic band approximation, as implemented in the \textit{SeeBand} software \cite{parzer2024seeband}. Within this framework, thermoelectric transport is modelled by numerically solving the respective Fermi transport integrals and summing up the contributions of the individual bands, assuming parallel conduction through two transport channels, \textit{i.e.}, one valence and one conduction band.

\begin{figure*}[tbh]
\newcommand{\setwidth}{0.45}
			\centering
			\hspace*{0cm}
			\includegraphics[width=0.8\textwidth]{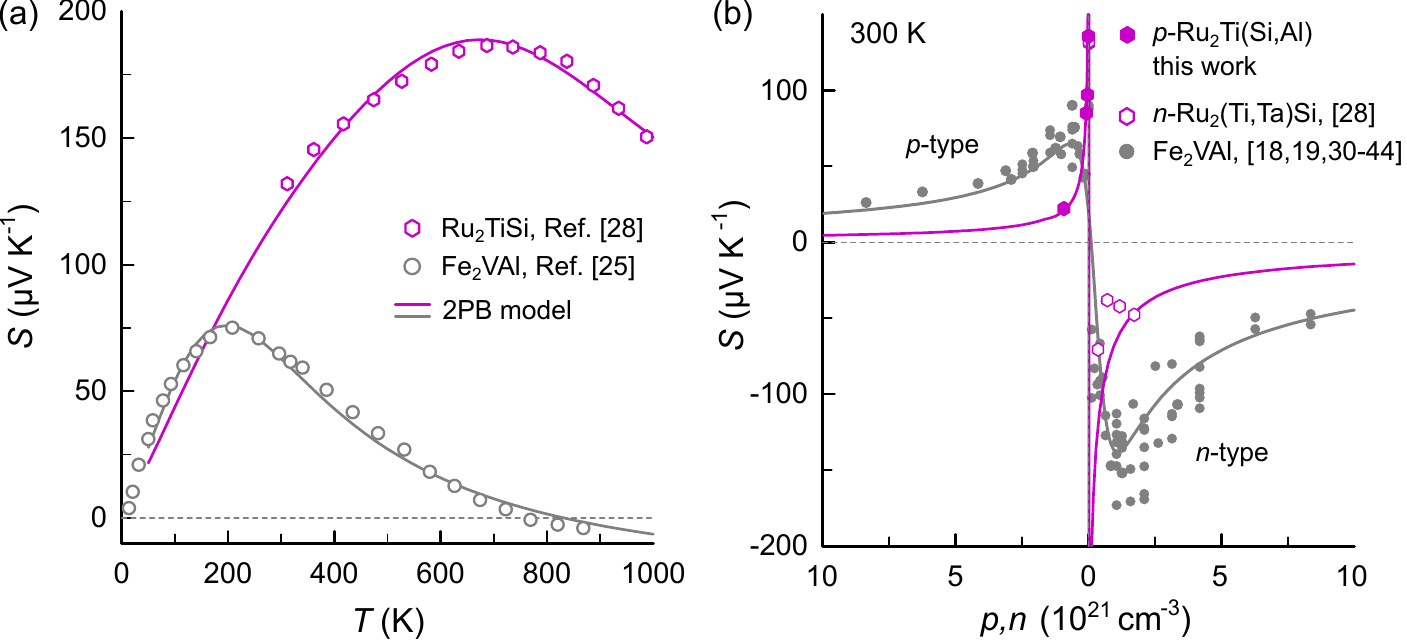}
	\caption{\textbf{Modelling of thermoelectric transport in \ce{Fe2VAl} and \ce{Ru2TiSi} full-Heuslers.} (a) Comparison of the temperature-dependent Seebeck coefficient of \ce{Fe2VAl} and \ce{Ru2TiSi}. Experimental data were taken from Refs.\,\cite{knapp2017impurity} and \cite{fujimoto2023enhanced}, respectively. Solid lines are least-squares fits employing a two-parabolic band model, which excellently captures the temperature-dependent $S(T)$ of both compounds in the entire temperature range. (b) Doping concentration-dependent Seebeck coefficient of $p$- and $n$-doped \ce{Fe2VAl} and \ce{Ru2TiSi} at room temperature. The Seebeck coefficient decreases much more rapidly as a function of the carrier concentration for the latter, indicating that doping is far more efficient in \ce{Ru2TiSi} compared to \ce{Fe2VAl}, owing to the much lighter and more dispersive bands; $p$-type \ce{Ru2TiSi_{1-x}Al_x}-based compounds (open symbols) were synthesized and investigated in this work. Filled purple symbols were taken from \cite{fujimoto2023enhanced} and filled grey symbols were taken from various literature studies \cite{skoug2009high,vasundhara2008electronic,
vasundhara2005low,nishino2006thermal,lue2007thermoelectric,
nishino2011development,mikami2013effect,nakayama2008thermoelectric,
terazawa2012effects,mikami2012thermoelectric,kato2001effect,
sandaiji2010off,lu2009thermoelectric,
hinterleitner2020stoichiometric,garmroudi2021boosting,
reumann2022thermoelectric,garmroudi2022large}.} 
	\label{Fig3}
\end{figure*}

\section{Results and Discussion}
\subsection{Ru$_2$TiSi versus Fe$_2$VAl}
In order to understand the thermoelectric properties and enhanced performance of \ce{Ru2TiSi}, a natural question would be what distinguishes it from the archetypal thermoelectric fH compound \ce{Fe2VAl}. Fig.\,\ref{Fig2}(a) shows the electronic density of states (DOS) of \ce{Ru2TiSi} and \ce{Fe2VAl} \footnote{Computational DOS data were taken from the Materials Project open web database \cite{jain2013commentary}.}. Both compounds display a deep, well pronounced pseudogap at the Fermi energy $E_\text{F}$, although the gap is significantly broader for \ce{Ru2TiSi}. Similar to \ce{Fe2VAl}, the DOS of \ce{Ru2TiSi} is characterized by sharp peaks (owing to the rather localized Ru-$4d$ states) rising next to both edges of the pseudogap, which results in a large differential DOS. Additionally, there are much more dispersive states (with a small DOS) dangling into the gap region, hereafter referred to as \textit{pseudogap states}. It was shown in previous studies that in \ce{Fe2VAl}, electronic transport is almost exclusively governed by these pseudogap states. Fig.\,\ref{Fig2}(b) shows the DOS around $E_\text{F}$ and a close-up of these pseudogap states. One immediately notices that the pseudogap states are much broader and the DOS much smaller for \ce{Ru2TiSi} than for \ce{Fe2VAl}, especially at $E<E_\text{F}$. This implies much lighter, more mobile charge carriers when $E_\text{F}$ is placed in the vicinity of these states. Moreover, the lower density of states effective mass $m^*_\text{DOS}$ suggests a much more efficient doping scenario for \ce{Ru2TiSi}. We illustrate this, by drawing and comparing the Fermi level of hole-doped \ce{Fe2VAl} (grey dashed line) and \ce{Ru2TiSi} (purple dashed line) for a hole doping concentration of 0.05 holes per formula unit, assuming rigid-band doping. This corresponds approximately to the concentration, for which $p$-type \ce{Fe2VAl} displays its optimal thermoelectric performance. It can be seen that, while for \ce{Fe2VAl}, this would place $E_\text{F}$ about 0.2\,eV below the valence band edge, $E_\text{F}$ is shifted by almost 0.6\,eV into the valence band for \ce{Ru2TiSi}. Hence, a significantly smaller number of holes has to be doped to reach optimal performance (compare also the integrated DOS in the inset of Fig.\,\ref{Fig2}(b)). Interestingly, the difference in $m^*_\text{DOS}$ appears less sizeable for the conduction band at $E>E_\text{F}$.

Fig.\,\ref{Fig3}(a,b) displays the temperature-dependent and doping concentration-dependent Seebeck coefficient ($S(T)$, $S(p,n)$), respectively. In Fig.\,\ref{Fig3}(a), we model experimetal $S(T)$ data of \ce{Fe2VAl} and \ce{Ru2TiSi} available in the literature \cite{knapp2017impurity,fujimoto2023enhanced}, by employing a two-parabolic band (2PB) least-squares fit model as implemented in the \textit{SeeBand} software package \cite{parzer2024seeband}. To fit the data, three independent fit parameters are adjusted: (i) The position of the Fermi level, (ii) the band gap/overlap $E_\text{g}$ and (iii) a weighting parameter $\epsilon_m = (N_\text{VB}m^*_\text{CB})/(N_\text{CB}m^*_\text{VB})$, which includes degeneracies $N_i$ and effective masses $m_i$ of the two bands ($i=\lbrace \text{VB},\,\text{CB} \rbrace$). $E_\text{F}$ determines the slope of $S(T)$ at low temperatures, $E_\text{g}$ the maximum Seebeck coefficient $S_\text{max}$ as well as the temperature of the maximum $T_\text{max}$, and $\epsilon$ dictates the sharpness of the maximum, that is, the tail of $S(T)$ in the bipolar regime at temperatures above $T_\text{max}$. Details regarding the modelling framework, which we previously successfully applied to a number of fH \cite{garmroudi2021boosting,garmroudi2022large} and skutterudite thermoelectric materials \cite{rogl2022understanding}, are described in Ref.\,\cite{parzer2024seeband}. Like for \ce{Fe2VAl}, excellent agreement with experimental data is also found for \ce{Ru2TiSi}. The somewhat similar slope of $S(T)$ at low temperatures for \ce{Fe2VAl} and the extrapolated curve of the 2PB model for \ce{Ru2TiSi} indicates a comparable position of $E_\text{F}$ with respect to the valence band edge. The most notable difference, however, is that $S_\text{max}$ is shifted towards much higher temperatures in \ce{Ru2TiSi} and reaches a very broad maximum of almost 200\,$\upmu$V\,K$^{-1}$, as opposed to only 70\,--\,80\,$\upmu$V\,K$^{-1}$ in \ce{Fe2VAl}. 

\begin{figure*}[t!]
\newcommand{\setwidth}{0.45}
			\centering
			\hspace*{0cm}
		\includegraphics[width=0.85\textwidth]{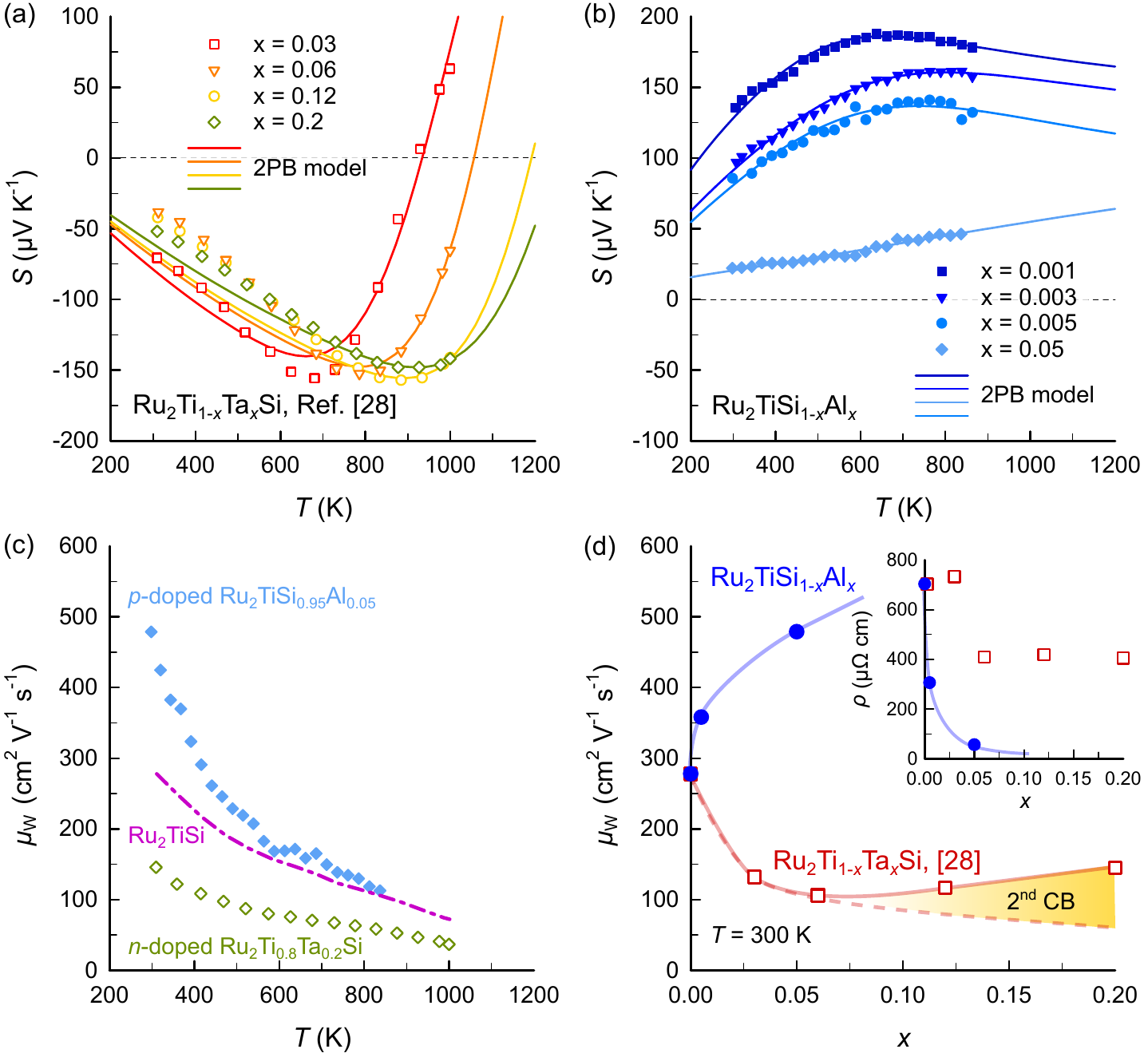}
	\caption{\textbf{Electron-hole band asymmetry in \ce{Ru2TiSi}.} Temperature-dependent Seebeck coefficient $S(T)$ of (a) $n$-type \ce{Ru2Ti_{1-x}Ta_xSi} from Ref.\,\cite{fujimoto2023enhanced} and (b) $p$-type \ce{Ru2TiSi_{1-x}Al_{1-x}} from this work. Solid lines are least-squares fits employing a two-parabolic band model. Notably, the maximum in $|S(T)|$ is much sharper for $n$-type \ce{Ru2Ti_{1-x}Ta_xSi} and even a sign reversal of the Seebeck coefficient takes place, once minority carriers from the valence band are activated. This implies that holes are much more mobile than carriers occupying the conduction band states in \ce{Ru2TiSi}. For $p$-type \ce{Ru2TiSi_{1-x}Al_{1-x}} the maximum in $|S(T)|$ is much broader and doping is much more efficient, that is, a substitution of only a few at.\% Al for Si results in a strong decrease of $S(T)$ as $E_\text{F}$ moves rapidly away from the band edge. (c) Weighted mobility comparison of heavily $n$-doped \ce{Ru2Ti_{0.8}Ta_{0.2}Si}, pristine \ce{Ru2TiSi} and heavily $p$-doped \ce{Ru2TiSi_{0.95}Al_{0.05}}. (d) shows the composition-dependent weighted mobility evaluated at 300\,K. Solid and dashed lines are a guide to the eye with the yellow area corresponding to an expected gain in $\mu_\text{W}$ arising from the contribution of a second conduction band. Most importantly, $\mu_\text{W}$ is much greater for $p$-type \ce{Ru2TiSi}, confirming its superiority over $n$-type \ce{Ru2TiSi}. Inset shows the corresponding composition-dependent resistivity.} 
	\label{Fig4}
\end{figure*}

Fig.\,\ref{Fig3}(b) shows the room-temperature Seebeck coefficient as a function of the hole ($p$) and electron ($n$) doping concentration, calculated from the VEC via $p,n=16$\,$\cdot$\text{VEC}/$a^3$, where $a^3$ denotes the cubic unit cell volume of \ce{Ru2TiSi} or \ce{Fe2VAl} and 16\,$\cdot$\text{VEC} represents the number of valence electrons in the unit cell. Each point refers to a different sample with a different VEC, which was varied through aliovalent element substitution. Data points were taken from numerous doping studies in the literature and $p$-type \ce{Ru2TiSi_{1-x}Al_x} compounds were synthesized and investigated in the course of this work. Solid lines are theoretical calculations employing a 2PB model. Anand \textit{et al.} previously showed that the $S(p,n)$ dependence of \ce{Fe2VAl} can be adequately described employing a 2PB model, assuming a small positive band gap of $E_\text{g}\approx 0.02$\,eV and valence and conduction band effective masses of $m^*_\text{VB}\approx 4.7\,m_e$ and $m^*_\text{CB}\approx 12.8\,m_e$, respectively \cite{anand2020thermoelectric}.

\begin{figure*}[t!]
\newcommand{\setwidth}{0.45}
			\centering
			\hspace*{0cm}
		\includegraphics[width=0.95\textwidth]{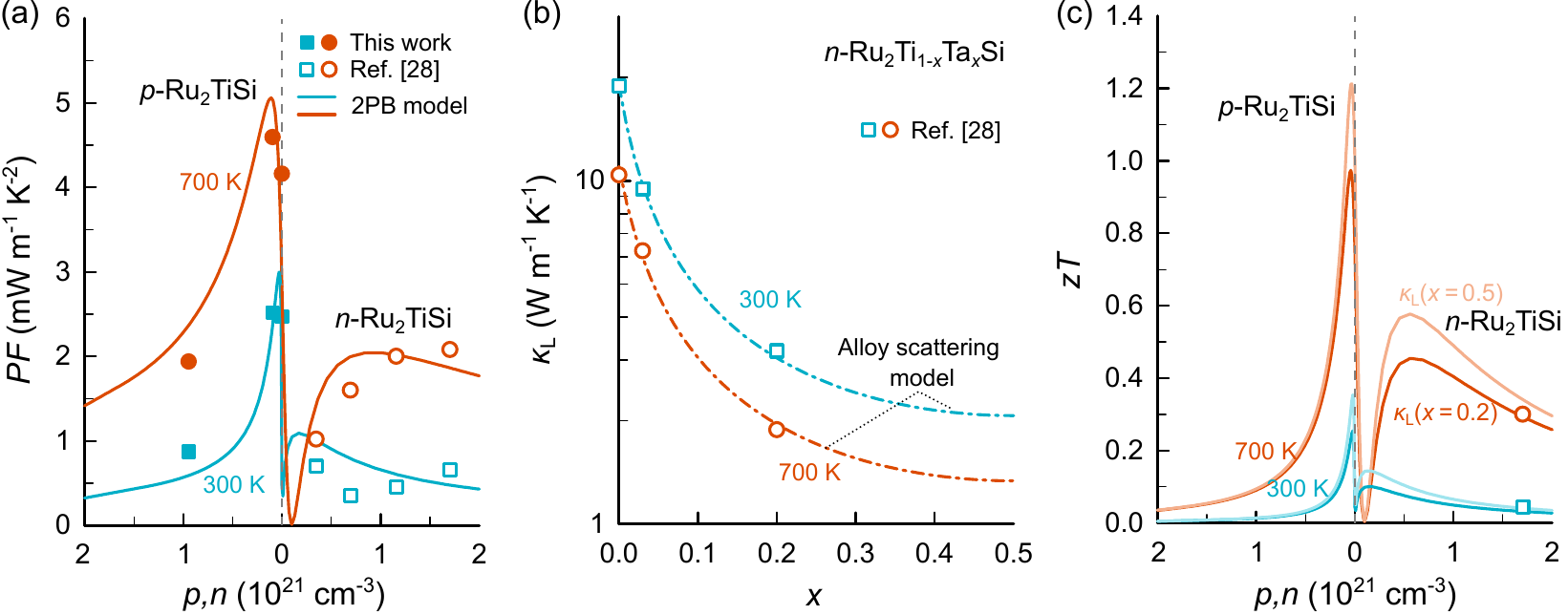}
	\caption{\textbf{Modelling and estimation of thermoelectric performance in \ce{Ru2TiSi}-based full-Heusler compounds.} (a) Doping concentration-dependent power factor $P\!F$ of \ce{Ru2TiSi} at 300\,K and 700\,K. Open and filled symbols are experimental data from this work and from Ref.\,\cite{fujimoto2023enhanced}, respectively. Solid lines are theoretical calculations using a two-parabolic band model. A more than two times larger $P\!F$ is predicted (and experimentally achieved) for $p$-type \ce{Ru2TiSi} due to the superior quality of the valence band structure. (b) Composition-dependent trend of lattice thermal conductivity for \ce{Ru2Ti_{1-x}Ta_xSi} from Ref.\,\cite{fujimoto2023enhanced}. Dashed-dotted lines were calculated employing a simple alloy scattering model \cite{anand2020thermoelectric}. The large mass and volume fluctuations imposed by the substitution with heavy 5$d$ elements effectively scatter high-frequency heat-carrying phonons. (c) Doping concentration-dependent $zT$ of \ce{Ru2TiSi} at 300\,K and 700\,K. The power factor and the electronic contribution to the thermal conductivity were calculated employing a two-parabolic band model. The lattice thermal conductivity was estimated from (b) under the reasonable assumption that the substitution with other 5$d$ elements, such as Hf/Ti, results in a similar suppression of $\kappa_\text{L}$. Hf substitution does not change the total number of valence electrons and, therefore, remains near the predicted maximum $zT$, exceeding $zT=1$ at 700\,K, especially in the case that a full solid-solution between \ce{Ru2TiSi} and \ce{Ru2HfSi} is possible.} 
	\label{Fig5}
\end{figure*}

Contrary to \ce{Fe2VAl}, we find from our temperature- and doping-dependent analysis of the Seebeck coefficient of \ce{Ru2TiSi} a much larger band gap of $E_\text{g}\approx 0.22$\,--\,0.24\,eV and much lighter effective masses $m^*_\text{VB}\approx 1.0\,m_e$ and $m^*_\text{CB}\approx 3.3\,m_e$. The effect of the latter is directly visible in the much more rapid decrease of $S$ as $p,n$ increase and is consistent with the smaller DOS and broader bandwidth of the pseudogap states of \ce{Ru2TiSi} shown in Fig.\,\ref{Fig2}(b). For $p$-type \ce{Ru2TiSi_{1-x}Al_x}, for instance, $S(300\,$K) decreases from 130\,$\upmu$V\,K$^{-1}$ in pristine \ce{Ru2TiSi} down to only 22\,$\upmu$V\,K$^{-1}$ for $x=0.05$. In this sense, the valence band electronic structure of \ce{Ru2TiSi} appears much more similar to chalcogenide semiconductors, such as \ce{Bi2Te3} and \ce{PbTe}, where the underlying DOS is composed of $s$- and $p$-like states with a larger bandwidth (low $m^*_\text{DOS}$), rather than half- and full-Heusler compounds, where usually a much larger doping concentration is required due to the more localized nature of the $d$ orbitals building up the DOS (high $m^*_\text{DOS}$). At this point, we also note that, interestingly, $S(n)$ first decreases and then increases with $n$ for $n$-type \ce{Ru2Ti_{1-x}Ta_xSi} reported in Ref.\,\cite{fujimoto2023enhanced}, instead of monotonically decreasing. While this might be within the margin of experimental uncertainties, it could also be an indication of the contribution of a second conduction band as $E_\text{F}$ is shifted further towards higher energies with increasing Ta substitution. Indeed, this would be consistent from a purely rigid-band-like shift of $E_\text{F}$, considering the DOS presented in Fig.\,\ref{Fig2}(b), which changes its slope around 0.4\,--\,0.5\,eV above $E_\text{F}$. 

\subsection{Detailed analysis of \textit{p}- and \textit{n}-doped \ce{Ru2TiSi}}
Next, we present a detailed analysis of the temperature-dependent transport properties of $n$-type \ce{Ru2Ti_{1-x}Ta_xSi} from Ref.\,\cite{fujimoto2023enhanced} and $p$-type \ce{Ru2TiSi_{1-x}Al_x} from this work. Fig.\,\ref{Fig4}(a) displays $S(T)$ data of the former, collected by Fujimoto \textit{et al.} in the temperature range 300\,--\,1000\,K. Least-squares fits using the 2PB model are again able to reasonably well reproduce all measured curves. Particularly striking is the distinct maximum  in $S(T)$ at 700\,K, followed by a sign reversal of $S(T)$ at around 900\,K for \ce{Ru2Ti_{0.97}Ta_{0.03}Si}. This is a direct signature of the strong electron--hole asymmetry and the much higher mobility of holes compared to the conduction band electrons. This is directly reflected in the weighting parameter $\epsilon_m$ derived from our fit. A large $\epsilon_m$ implies that either a much larger degenerate set of hole pockets (compared to the electron pockets) contributes to the transport properties and/or that conduction band electrons are much heavier compared to the hole-type carriers. An extraordinarily large value of $\epsilon_m\approx 60$ is found for \ce{Ru2Ti_{0.97}Ta_{0.03}Si} (and even larger values are found for higher Ta concentrations, see Appendix C), which is especially remarkable, considering that $N_\text{VB} = N_\text{CB} = 3$ is derived from band structure calculations provided in the Materials Project open web database \footnote{The valence band at $\Gamma$ is triply degenerate and the conduction band comprises sixfold generate half-pockets at X, thus, also equaling $N=3$.}. We attribute this extreme electron-hole asymmetry to the very dispersive valence band and the much more localized and heavy conduction band states, approximately 0.5\,eV above $E_\text{F}$ (see Fig.\,\ref{Fig2}(b)). These states become likely important at elevated temperatures and higher doping concentrations, which explains why a much smaller band asymmetry $\epsilon_m\approx 3.3$ is derived from the analysis of the doping-dependent Seebeck coefficient of \ce{Ru2TiSi}-based compounds at 300\,K, shown in Fig.\,\ref{Fig3}(b). Nonetheless, this band asymmetry is likely critical for the thermoelectric performance of \ce{Ru2TiSi}, which peaks at much higher temperatures $T>300\,$K.

The notion of much lighter holes compared to the conduction electrons is also confirmed when examining the temperature-dependent Seebeck coefficient of $p$-type \ce{Ru2TiSi_{1-x}Al_x} in Fig.\,\ref{Fig4}(b). Both, the experimental data collected in this study and the 2PB model extrapolating towards higher temperatures, show a much broader maximum of $S(T)$, almost saturating at high temperatures. This demonstrates that the weighted contribution of the minority carriers, activated at high temperatures is small. Fig.\,\ref{Fig4}(c) and (d) display the temperature- and composition-dependent weighted mobility $\mu_\text{W}$, calculated via the scheme presented in Ref.\,\cite{snyder2020weighted}. Despite the limitations of such an analysis with respect to $\mu_\text{W}$ owing to (i) bipolar transport and (ii) the rather complex band structure, particularly concerning the conduction band states, it is nonetheless obvious that $\mu_\text{W}$ is several times larger for $p$-doped \ce{Ru2TiSi}, as opposed to the $n$-doped compounds reported previously \cite{fujimoto2023enhanced}.

\subsection{Assessing optimal thermoelectric performance}
To evaluate the optimal performance of $p$- and $n$-type Ru$_2$TiSi, we modeled the composition-dependent power factor and thermal conductivity and the maximum $zT$ was estimated based on these theoretical results. Fig.\,\ref{Fig5}(a) shows the carrier doping-dependent $P\!F$ at 300\,K and 700\,K. Despite some limitations, our 2PB model traces the trend of the experimental data fairly well and correctly reproduces the enhanced performance of the $p$-type compounds. The electronic part of the thermal conductivity was calculated via the Wiedemann-Franz law, following the same procedure described in Ref.\,\cite{anand2020thermoelectric}. The lattice thermal conductivity $\kappa_\text{L}$, shown in Fig.\,\ref{Fig5}(b), can be described by a simple alloy scattering model \cite{anand2020thermoelectric}, which considers two primary scattering contributions, namely, (i) Umklapp phonon--phonon scattering and (ii) scattering of high-frequency phonons with point defects introduced by the random substitution of other atoms in the compound. The latter depends on mass and volume fluctuations and is especially significant when substituting heavy 5$d$ elements, \textit{e.g.} Ta instead of Ti. Following this argument, one may assume that other 5$d$ elements with similar atomic mass and size, such as Hf, would have a similar effect and composition dependence of $\kappa_\text{L}$, which is indeed what is observed in \ce{Fe2VAl} fH compounds as well \cite{alleno2018review}. According to the Materials Project database, \ce{Ru2HfSi} is also a theoretically predicted stable Heusler compound, as is \ce{Ru2ZrSi}, both with a very similar electronic structure to \ce{Ru2TiSi}. This suggests that alloying of \ce{Ru2TiSi} with \ce{Ru2ZrSi} and \ce{Ru2HfSi} could be possible, consequently leading to reduced $\kappa_\text{L}$, and hence increased $zT$. Since the valence band states of full-Heusler compounds, such as \ce{Fe2VAl} and \ce{Ru2TiSi}, are almost exclusively governed by the $X$ atoms, introducing disorder at the $Y$ site hardly affects the electronic transport properties and retains high values of the weighted mobility \cite{anand2020thermoelectric}. On the other hand, the pseudogap states of the conduction band have $Y$-$e_g$ orbital character, which results in a strong tradeoff between $\kappa_\text{L}$ and $\mu_\text{W}$. Thus, substituting heavy 5$d$ elements at the Ti site would be especially promising for $p$-type \ce{Ru2TiSi}, where $E_\text{F}$ is located in the Ru-$t_{2g}$ valence bands.

Fig.\,\ref{Fig5}(c) shows the theoretical prediction of the doping-dependent $zT$ for an alloy of \ce{Ru2TiSi} and \ce{Ru2HfSi} from our 2PB model, assuming similar electronic transport. Darker colors refer to a 20\,\% Ti-Hf substitution, while the lighter colors refer to a 50\,\% alloy, minimizing the lattice thermal conductivity of the compound. For the \ce{Ru2Ti_{0.5}Hf_{0.5}Si} alloy, our calculations predict a high maximum $zT = 1-1.2$ at 700\,K for optimal doping, which motivates experimental exploration of this virginal material platform and showcases that not only half-Heusler but also full-Heusler compounds bear the potential for competitive thermoelectric performance. 

\section{Conclusions}
Summarizing, we have investigated thermoelectric transport of \ce{Ru2TiSi}-based full-Heusler compounds and compared their transport properties to those of \ce{Fe2VAl}. A two-parabolic band model accurately captures the temperature- and doping-dependent thermoelectric transport properties of \ce{Ru2TiSi}. The resulting effective band structure underlies that the valence band electronic structure displays much greater potential for realizing high thermoelectric performance compared to $p$-type \ce{Fe2VAl}. Moreover, we predict that $p$-type \ce{Ru2TiSi} would outperform $n$-type compounds, studied previously, by a factor of 2\,--\,3, potentially realizing $zT>1$ at 700\,K upon appropriate reduction of the lattice thermal conductivity, \textit{e.g.}, by isovalent substitution with Zr or Hf at the Ti site. Our work encourages further investigation of the vast phase space of full-Heusler next to half-Heusler compounds as thermoelectric materials.

\section*{Acknowledgments}
F.G., M.P., E.B. and T.M. were financially supported by the Japan Science and Technology Agency (JST) program MIRAI, JPMJMI19A1. A.P. acknowledges support from OeAD WTZ (Project CZ 08/2023).

\section*{Appendix A: Structural Properties}
\begin{figure}[t!]
\newcommand{\setwidth}{0.45}
			\centering
			\hspace*{-0.0cm}
		\includegraphics[width=0.45\textwidth]{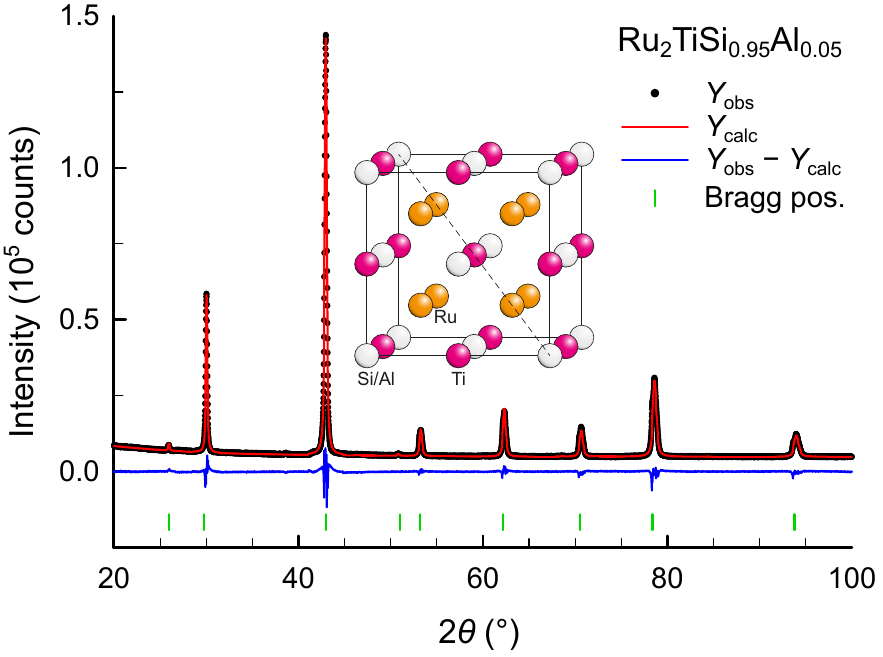}
	\caption{X-ray powder diffraction pattern of phase-pure $p$-type \ce{Ru2TiSi_{0.95}Al_{0.05}} alongside Rietveld refinement. Inset shows the full-Heusler crystal structure.} 
	\label{Fig6}
\end{figure}

The structural properties and phase purity of $p$-type \ce{Ru2TiSi_{1-x}Al_x} Heusler compounds, synthesized and investigated in this work, have been studied using x-ray powder diffraction (XRPD) making use of a commercially available diffractometer (AERIS by PANalytical). Cu-K$\alpha$ radiation has been used and measurements were conducted in a Bragg-Brentano geometry. An exemplary XRPD pattern for the sample with the largest amount of Al substitution, \textit{i.e.} \ce{Ru2TiSi_{0.95}Al_{0.05}}, is shown in Fig.\,\ref{Fig6} together with Rietveld refinements, which have been performed using the program PowderCell. The powder pattern displays a single full-Heusler (L$2_1$) phase with a lattice parameter $a\approx 0.5967$\,nm derived from the refinement. The room-temperature lattice parameter is only about 0.14\,\% larger than that of pristine \ce{Ru2TiSi} \cite{fujimoto2023enhanced}, which aligns with expectations since the atomic radii of Si ($r_\text{Si}\approx 111$\,pm) and Al ($r_\text{Si}\approx 125$\,pm) are quite similar.

\begin{figure*}[t!]
\newcommand{\setwidth}{0.45}
			\centering
			\hspace*{0cm}
		\includegraphics[width=0.8\textwidth]{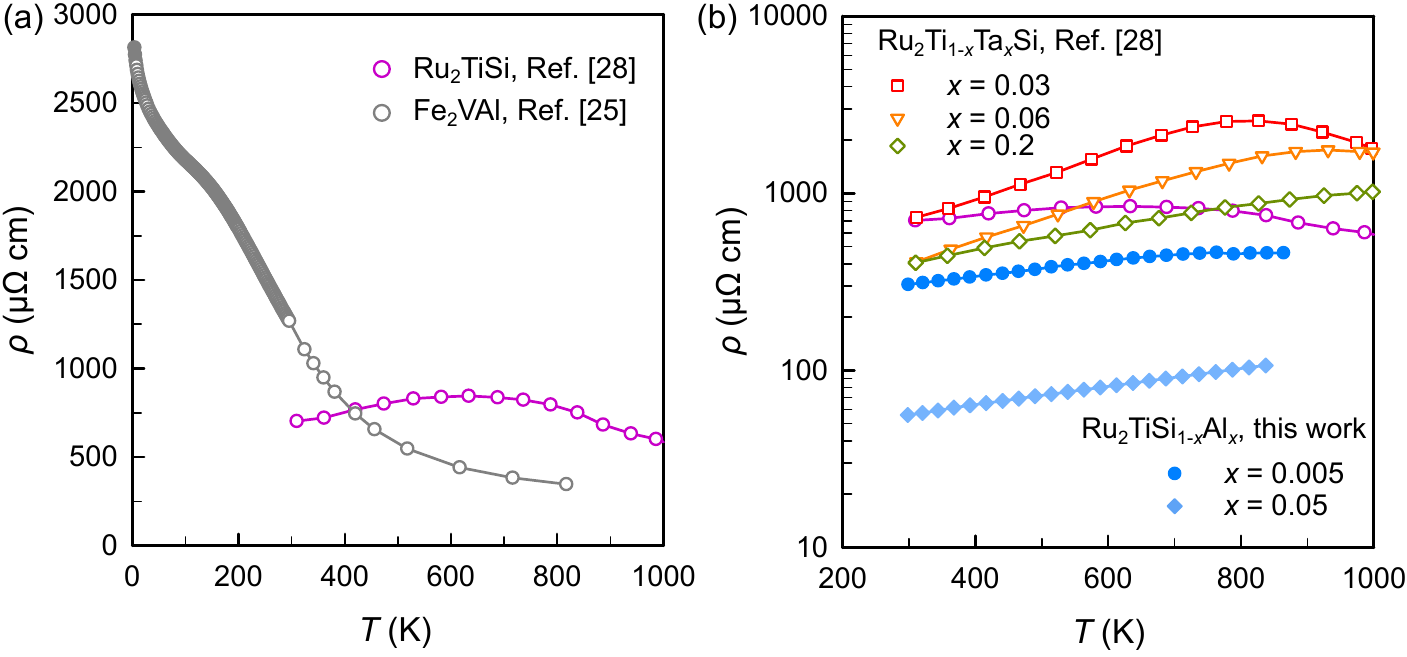}
	\caption{\textbf{Temperature-dependent electrical resistivity of \ce{Ru2TiSi} Heusler compounds.} (a) Temperature-dependent electrical resistivity of pristine \ce{Ru2TiSi} \cite{fujimoto2023enhanced} compared to that of pristine \ce{Fe2VAl} \cite{knapp2017impurity}. Although \ce{Ru2TiSi} has a larger band gap, the electrical resistivity, especially at low temperatures, is lower than that of \ce{Fe2VAl}, highlighting that $E_\text{F}$ is intrinsically doped in the valence band, which due to its lower effective mass enables a higher carrier mobility and lower resistivity for \ce{Ru2TiSi}. (b) Temperature-dependent resistivity for different $n$- and $p$-doped \ce{Ru2TiSi} samples from Ref.\,\cite{fujimoto2023enhanced} and this work, respectively. It is evident that for $p$-type \ce{Ru2TiSi_{1-x}Al_x}, $\rho(T)$ decreases much more rapidly as a function of the doping concentration.} 
	\label{Fig7}
\end{figure*}

\begin{figure*}[t!]
\newcommand{\setwidth}{0.45}
			\centering
			\hspace*{0cm}
		\includegraphics[width=0.8\textwidth]{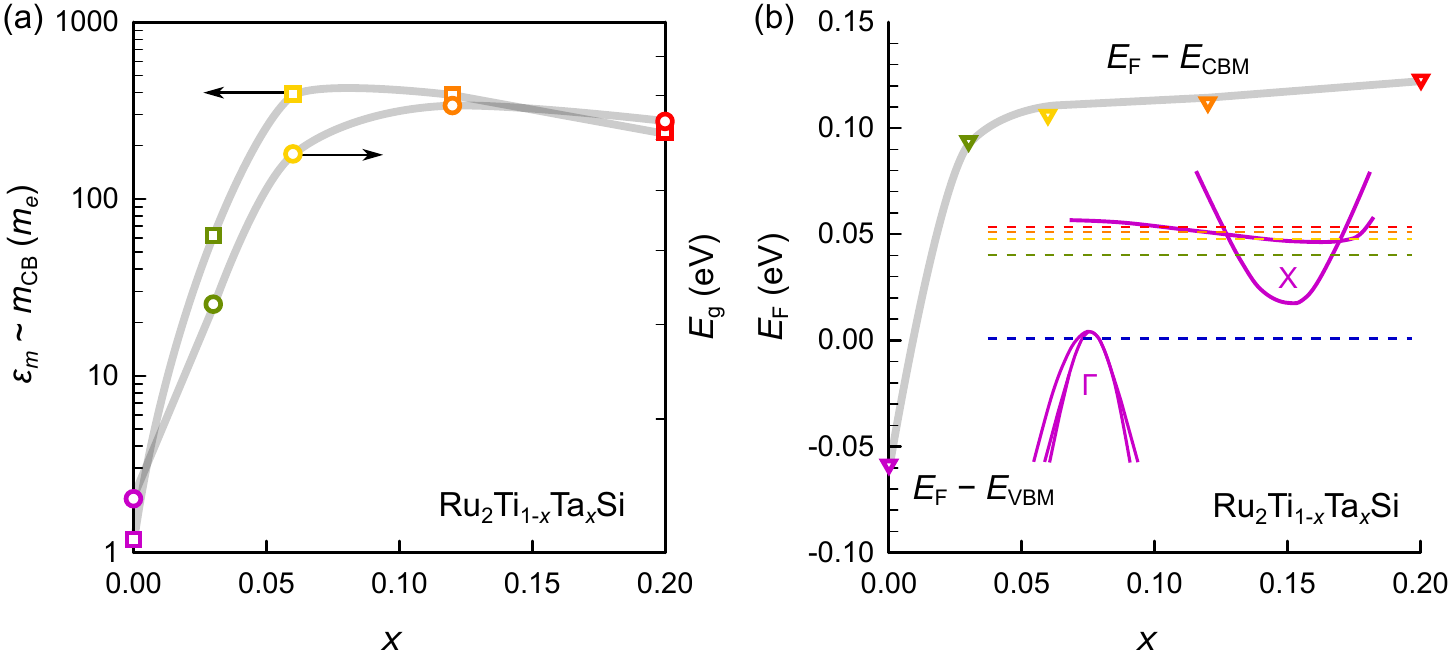}
	\caption{\textbf{Evolution of the electronic structure in $n$-doped \ce{Ru2Ti_{1-x}Ta_xSi}, extracted from fits of the temperature-dependent Seebeck coefficient.} (a) Concentration-dependent weighting parameter $\epsilon_m$ (left axis) and band gap $E_\text{g}$ (right axis) of $n$-type \ce{Ru2Ti_{1-x}Ta_xSi}. The former corresponds to the effective mass of the conduction band, given that $m_\text{VB}\approx 1\,m_e$ and $N_\text{VB} = N_\text{CB}$. Both $\epsilon_m \sim m_\text{CB}$ and $E_\text{g}$ extracted from our 2PB fits increase with $x$ as $E_\text{F}$ is shifted further towards the flat band, located 0.75\,eV above the top of the valence bands. (b) Doping level of $n$-type \ce{Ru2Ti_{1-x}Ta_xSi}. $E_\text{F}$ abruptly jumps from the valence towards the conduction band but saturates with increasing $x$ due to the high DOS of the flat-band states.} 
	\label{Fig8}
\end{figure*}

\section*{Appendix B: Electrical resistivity}
Fig.\,\ref{Fig7} summarizes the temperature-dependent electrical resistivity $\rho(T)$ of the \ce{Ru2TiSi} system. First, $\rho(T)$ of pristine \ce{Ru2TiSi} from Ref.\,\cite{fujimoto2023enhanced} is compared with $\rho(T)$ of pristine \ce{Fe2VAl} from Ref.\,\cite{knapp2017impurity} in Fig.\,\ref{Fig7}(a). Despite the larger band gap, evident from the larger Seebeck coefficient of \ce{Ru2TiSi}, the resistivity at low temperatures is actually lower compared to \ce{Fe2VAl}. We attribute the fact, that \ce{Ru2TiSi} is intrinsically doped, with $E_\text{F}$ located about 0.06\,eV below the valence band edge. Thus, at low temperatures $\rho(T)$ should show metallic-like behavior, as is indeed observed. Fujimoto \textit{et al.} reported a carrier mobility of around 100\,cm$^2$\,V$^{-1}$\,s$^{-1}$ for pristine \ce{Ru2TiSi} \cite{fujimoto2023enhanced}, about an order of magnitude larger than that of \ce{Fe2VAl}. The carrier mobility in undoped \ce{Fe2VAl} is not only smaller due to less dispersive bands at $E_\text{F}$, but also hampered by pivotal carrier scattering off localized in-gap impurity states arising from intrinsic Fe/V and Fe/Al exchange antisite defects \cite{garmroudi2023pivotal}. It is possible that the formation of such antisite defects, involving the Ru sublattice is suppressed in \ce{Ru2TiSi} due to the larger atomic size mismatch between Ru and Ti/Si, as opposed to Fe versus V/Al.

Fig.\,\ref{Fig7}(b) shows $\rho(T)$ for various $n$- and $p$-doped \ce{Ru2TiSi}-based fHs from Ref.\,\cite{fujimoto2023enhanced} and this work, respectively. It is evident, that for $p$-type compounds, $\rho(T)$ decreases extremely quickly, again reflecting the dispersive nature of the valence band electronic structure. For \ce{Ru2TiSi_{1-x}Al_x}, an Al substitution of only $x=0.05$, yields a very strong decrease of the room-temperature resistivity down to only $56$\,$\upmu\Omega$\,cm, which is comparable to that of ordinary metals.

\section*{Appendix C: Fit parameters}
\begin{table}[b!]
\centering
\caption{Obtained fit parameters from modelling the temperature-dependent Seebeck coefficient of \ce{Ru2Ti_{1-x}Ta_{x}Si} and \ce{Ru2TiSi_{1-x}Al_x}. The values of Fermi energy are given with respect to the valence/conduction band edge for $p$- and $n$-type samples respectively.}
\vspace{0.1cm}
\begin{tabular}{c c c c c c c c c c c}

\hline
\hline
\noalign{\vskip 0.2cm}
\vspace{0.2cm}

Sample \,& &\, $x$ \,& & \,\,\,\,$\epsilon_m$ \,& & \,$E_\text{g}$ (eV) \,& & \,$E_\text{F}$ (eV)  \\
\hline
\\ 
$n$-type \ce{Ru2Ti_{1-x}Ta_{x}Si} 
\,& & 0 & & \,\,\,\,1.2 & & \,0.24 & &  $-0.06$\\ 

\,& & 0.03 & & \,\,\,\,62 & & \,\,\,\,0.42 & & 0.09\\ 

\,& & 0.06 & & \,\,\,\,392 & &\,\,\, 0.67 & & 0.11\\ 

\,& & 0.12 & & \,\,\,\,385 & &\,\,\, 0.78 & & 0.11\\

\,& & 0.20 & & \,\,\,\,239 & &\,\,\, 0.74 & & 0.12\\  
 \\
$p$-type \ce{Ru2TiSi_{1-x}Al_x} 
\,& & 0 & & \,\,\,\,1.2 & & \,0.24 & &  $-0.06$\\
 
\,& & 0.001 & & \,\,\,\,3.4 & & \,\,\,\,\,0.18 & & $-0.05$\\ 

\,& & 0.003 & & \,\,\,\,2.1 & & \,\,\,\,\,0.16 & & $-0.08$\\ 

\,& & 0.005 & & \,\,\,\,1.5 & & \,\,\,\,\,0.11 & & $-0.09$\\ 

\,& & 0.05 & & \,\,\,\,- & & \,\,\,\,\,- & & $-0.26$\\ 
\noalign{\vskip 0.3cm}
\hline
\hline
\end{tabular} 
\\ \vspace{0.1cm}
\label{Fit_parameters}
\end{table}
\autoref{Fit_parameters} lists the obtained fit parameters from our analysis of the temperature-dependent Seebeck coefficient of $n$-type \ce{Ru2Ti_{1-x}Ta_{x}Si} from Ref.\,\cite{fujimoto2023enhanced} and $p$-type \ce{Ru2TiSi_{1-x}Al_x} from this work. There are three independent fit parameters that can be extracted to develop an effective band structure model. The first parameter, $\epsilon_m$ serves as a weighting parameter, which represents the weighted contribution between the two bands. The larger $\epsilon_m$, the smaller the weighted contribution of the conduction band carriers. Since $m_\text{VB}\approx 1\,m_e$ could be derived from the carrier concentration dependence of $S$ and because $N_\text{VB}= N_\text{CB}\approx 3$, $\epsilon_m$ can be considered the effective mass of the conduction band electrons $\epsilon_m \sim m_\text{CB}$. Taking a look at the values in \autoref{Fit_parameters}, it is clear that $m_\text{CB}$ dramatically increases with Ta substitution, and sort of saturates at very large values of several hundred times the free electron mass (left axis in Fig.\,\ref{Fig8}(a)). This aligns with the notion of a second, much heavier conduction band, further above $E_\text{F}$, which is also observed in $\ce{Fe2VAl}$ and similar Heusler compounds and whose origin has been extensively discussed by Bilc \textit{et al.} \cite{bilc2015low}. Similarly, the band gap derived from  our 2PB model seemingly increases with $x$ and saturates around $0.7-0.8$\,eV (right axis in Fig.\,\ref{Fig8}(a)). We interpret this as the position of the second conduction band minimum with respect to the valence band top. The Fermi level $E_\text{F}$, which is given with respect to the top of the valence band for $p$-type and with respect to the bottom of the conduction band for $n$-type samples, rapidly jumps from the top of the valence band and is shifted into the conduction band states. As $x$ increases further, however, $E_\text{F}$ almost saturates and is seemingly pinned in the conduction band (see Fig.\,\ref{Fig8}(b)), which is another indirect proof of the flat and heavy band and its associated high DOS that prevents efficient doping (shifts of $E_\text{F}$). A schematic of the effective band structure expected for \ce{Ru2TiSi}, the respective energy gaps and position of the Fermi level is presented in Fig.\,\ref{Fig8}(b).

Summarizing, there is conclusive indirect evidence from various fit parameters obtained by modelling the temperature-dependent Seebeck coefficient, that \ce{Ru2TiSi} is a narrow-gap semiconductor with dispersive valence and conduction band states and significant electron-hole asymmetry arising from another conduction band hosting significantly (orders of magnitude) heavier charge carriers, especially when $E_\text{F}$ is shifted further into the conduction band via $n$-type dopants and/or at high temperature where the Fermi-Dirac distribution broadens and excites states further away from $E_\text{F}$.

%

\end{document}